\documentclass[namedreferences]{solarphysics}
\usepackage[optionalrh]{spr-sola-addons} 
\usepackage{graphicx}        

\usepackage{color}           
\usepackage{sidecap}
\usepackage{xurl}

\usepackage{setspace}
\usepackage[T1]{fontenc}

\newcommand{\kms}{km s$^{-1}$}

\chardef\us=`\_

\usepackage[bookmarks = true,colorlinks = true,linkcolor = blue,urlcolor = blue,citecolor = blue,bookmarks = true,linktocpage = true,hyperindex = true,breaklinks]{hyperref}

\begin{document}

\begin{article}
\begin{opening}

\title{ Extreme-Ultraviolet Wave and Accompanying Loop Oscillations}

\author[addressref={aff1},corref,email={setiapooja.ps@gmail.com}]{\inits{P.}\fnm{Pooja}~\lnm{Devi}\orcid{0000-0003-0713-0329}}
\author[addressref={aff1}]{\inits{R.}\fnm{Ramesh}~\lnm{Chandra}\orcid{0000-0002-3518-5856}}
\author[addressref={aff2}]{\inits{A.}\fnm{Arun Kumar}~ \lnm{Awasthi}\orcid{0000-0003-1948-1548}} \author[addressref={aff3,aff4,aff5}]{\inits{B.}\fnm{Brigitte}~\lnm{Schmieder}\orcid{0000-0003-3364-9183}}
\author[addressref={aff6,aff7}]{\inits{R.}\fnm{Reetika}~\lnm{Joshi}\orcid{0000-0003-0020-5754}}

\address[id=aff1]{Department of Physics, DSB Campus, Kumaun University, Nainital 263 001, India}
\address[id=aff2]{Space Research Centre, Polish Academy of Sciences, Bartycka 18A, 00-716 Warsaw, Poland}
\address[id=aff3]{LESIA, Observatoire de Paris,  Universit\'e PSL, CNRS, Sorbonne Universit\'e,  Universit\'e de Paris, 5 place Janssen, 92290  Meudon Principal Cedex, France}
\address[id=aff4]{Centre for mathematical Plasma Astrophysics, Dept. of Mathematics, KU Leuven, 3001 Leuven, Belgium}
\address[id=aff5]{SUPA, School of Physics and Astronomy, University of Glasgow, Glasgow G12 8QQ, Glasgow, UK}
\address[id=aff6]{Institute of Theoretical Astrophysics, University of Oslo, P.O. Box 1029 Blindern, N-0315 Oslo, Norway}
\address[id=aff7]{Rosseland Centre for Solar Physics, University of Oslo, P.O. Box 1029 Blindern, N-0315 Oslo, Norway}
\runningauthor{Devi et al.}
\runningtitle{Filament eruption and EUV wave}

\begin{abstract}

We present the observations of an extreme-ultraviolet (EUV) wave, which originated from the active region (AR) NOAA 12887 on 28 October 2021 and  its impact  on neighbouring loops.
The event was observed by the {\it Atmospheric Imaging Assembly (AIA)} on board the {\it Solar Dynamics Observatory (SDO)} satellite at various wavebands and by the {\it Solar TErrestrial RElations Observatory-Ahead (STEREO--A)} with its {\it Extreme-Ultraviolet Imager (EUVI)} and COR1 instruments with a different view angle than SDO. We show that the EUV wave event consists of  several waves as well as non-wave phenomena.
The wave components include: the fast--mode part of the EUV wave event, creation of oscillations in nearby loops, and the appearance of wave trains. 
The non-wave component consists of stationary fronts.
We analyze selected oscillating loops and find that the periods of these oscillations range from 230 -- 549 s. Further, we compute the density ratio inside and outside the loops and the magnetic field strength. The computed density ratio and magnetic field are found in the range of 1.08 -- 2.92 and 5.75 -- 8.79 G, respectively.
Finally, by combining SDO and STEREO-A observations, we find that the observed EUV wave component propagates ahead of the CME leading edge.
\end{abstract}
\end{opening}


\section{Introduction}
     \label{S-Introduction} 

Extreme-ultraviolet (EUV) waves are defined as large propagating fronts on the whole solar surface with a speed ranging from a few 100 \kms~to multiple of 1000 \kms~\citep[see the review by][]{Warmuth2015}. Because of their clear observation in EUV wavelengths, they are named ``EUV waves''.
However, EUV waves are also known by several other names, for example, EIT waves (where EIT stands for Extreme-ultraviolet Imaging Telescope), coronal waves, coronal propagating fronts (CPFs), and so on \citep{Wang2000, Wu2001, Nitta2013, Schrijver2011}.
They are mostly associated with solar filament eruptions, flares, and coronal mass ejections (CMEs) \citep{Wang2000, Chen2011, Chandra2016,
Chandra2018}. After the increased high spatial and temporal resolution data sets of the Solar Dynamics Observatory \citep[SDO:][]{Pesnell2012}, the association between the successful and/or partial filament eruption and the EUV wave became stronger
\citep{Chen2011, Asai2012, Chandra2016, Chen2017, Zong2017, Chandra2021}. 
This is due to their connection with CMEs \citep{Schmieder2013, Chandra2011, Seki2021}.
 
In the studies of EUV waves, several features are reported. These features include: wave, non-wave components, stationary fronts, \citep{Chen2002, Chen2011, Schrijver2011, Asai2012, Chandra2021}. Apart from these features, the reflection \citep{Long2008, Veronig2008, Gopalswamy2009, Schmidt2010, Shen2013}, refraction \citep{Thompson2000, Hudson2003, Patsourakos2009, Kienreich2009, Liu2012, Shen2013}, mode conversion \citep{Chen2002, Chen2005, Chandra2018}, and the nearby loop and filament oscillations \citep{Ballai2005, Guo2015, Fulara2019, Shen2019, Devi2022} were also reported. An interesting feature of EUV waves, called wave trains, has been also observed \citep{Liu2012, Shen2019, Zhou2021, Zhou2022, Shen2022}.  
As far as the interpretations of EUV wave trains, \inlinecite{Liu2012} proposed that the wave trains are driven by downward and lateral compression of CMEs. On the other hand, \inlinecite{Shen2019} concluded that the unwinding motion of the filament helical structures can also produce EUV wave trains. Due to limited observational case studies, their driving mechanism is not well understood and needs to be studied in detail and comprehensively. 

To explain the nature of the EUV waves, several models were proposed. They 
are mainly divided into three categories. The first is the MHD wave model \citep{Wu2001, Chen2005, Schmidt2010}, the second is related to the non-wave or pseudo wave model \citep{Chen2005, Delannee2007}, and the 
third category of the EUV wave models is the hybrid model proposed by \inlinecite{Chen2002}. 
 Further according to wave models, the waves can be classified as fast-mode MHD waves \citep{Thompson1998,  Wang2000,  Wu2001, Patsourakos2009, Schmidt2010}, slow-mode solitons \citep{Wills-Davey2007}, and magnetoacoustic surface gravity waves \citep{Ballai2011}. According to the fast-mode MHD wave model, they can be 
  driven by shock waves and are known as magnetoacoustic waves. These shock wave can be formed  by either  a three-dimensional piston driven by a CME expansion or by a pressure pulse triggered by a solar flare. In the fast-mode wave model, the magnetic and  gas pressure act as restoring forces. Solitons are non-linear waves where the non-linear steepening is compensated by dispersive effects \citep{Warmuth2015}.  
According to the non-wave/pseudo wave model, the pseudo waves are generated by the restructuring of the coronal magnetic configuration due to the expanding flux rope or CME. As a result of this reconfiguration, the propagating brightening is observed. 

One of the interesting features of the EUV waves is the creation of oscillations, when they  encounter  coronal loops \citep{Aschwanden1999, Zheng2013, Zimovets2015, Fulara2019, Nakariakov2021} or filaments/prominences \citep{Okamoto2004, Isobe2006, Shen2019, Devi2022}. These oscillations later show a decay in amplitude \citep{Nakariakov1999, Aschwanden2002, Jing2006, Vrsnak2007, Arregui2008, Hershaw2011}. The possible damping mechanism could be due to  resonant absorption \citep{Goossens2002, Ruderman2002} or  cooling of the plasma \citep{Morton2009, Morton2010, Morton2010a}. However, there are reports
where the amplitude remains constant \citep{Anfinogentov2013, Anfinogentov2015}. Since coronal loop oscillations are 
important phenomena, people are using them to study coronal seismology to derive coronal physical parameters such as the coronal magnetic field, density ratio, radius of curvature, etc. Therefore, it is important and useful to compare the coronal parameters obtained from the seismology and theoretical models \citep{Edwin1983, Roberts1984, Andries2009, Su2018}.  

There are several questions about the nature of EUV waves including:
How  are the EUV waves generated by flares or filament eruptions? Why are there multiple wavefronts observed in some EUV waves? What are the characteristics of oscillating loops hit by  EUV waves? To address these questions, 
we analyze observed features in EUV, and the kink oscillations in nearby loops triggered by the interaction between these loops and the EUV wave on 28 October 2021 originated from NOAA active region (AR) 12887. This  article is structured as follows: Section \ref{S-Results} presents the observations and analysis. We discuss our results in Section \ref{S-discussion}. Finally, Section \ref{S-Conclusion} is devoted to the conclusion of the study.
\\

\section{Observations} 
      \label{S-Results}  
      
     \subsection{Instruments} 
  
For the present study, we  use the data from the Atmospheric Imaging Assembly (AIA: \opencite{Lemen2012}) on board the Solar Dynamics Observatory (SDO: \opencite{Pesnell2012}), which observes the full Sun in seven EUV (94, 131, 171, 193, 211, 304, and 335 \AA), two ultraviolet (UV) (1600 and 1700 \AA), and one white light (4500 \AA) channels. 
The pixel resolution and cadence of EUV AIA observations are 0.6$''$ and 12 sec, respectively. In this study, we use all EUV channels except 304 \AA. 
For a different viewpoint of the event, we use data from  the Solar TErrestrial RElations Observatory-Ahead \citep[STEREO-A:][]{Kaiser2008} EUVI telescope in 195 \AA~wavelength, which observes the solar corona until 1.7 R$_{sun}$. EUVI is a part of the Sun Earth Connection Coronal and Heliospheric Investigation \citep[SECCHI;][]{Howard2008} instrument.
The pixel and temporal resolution of the EUVI data in this wavelength are 1.6$''$ and  2.5 min, respectively.
For the associated CME, the data from the Large Angle Spectroscopic Coronagraph \citep[LASCO;][]{Bruckner1995} on board the Solar and Heliospheric Observatory \citep[SOHO;][]{Domingo1995} and the STEREO-A COR1 \citep{Thompson2003} instruments are investigated. LASCO has two coronagraphs C2 and C3, whose field of view range from 2 to 6 R$_{sun}$ and 3.7 to 30 R$_{sun}$, respectively. The STEREO COR1 observes the solar corona from 1.5 to 4 R$_{sun}$.
     
\subsection {Overview of the event}
     
One of the large solar flare of the ongoing Solar Cycle 25 was observed on 28 October 2021 in  NOAA AR 12887, located at S26W07.
Its GOES classification is X1.0 class. The flare started at approximately 15:17 UT, reached its maximum around 15:35 UT, and progressed  into a decline phase afterward. It was a long duration event associated with two flare ribbons. The flare produced emission in almost all layers of the solar atmosphere and was associated with several peculiar features, which include: filament eruptions, EUV waves, Moreton waves, CMEs, ground-level enhancement (GLE), etc. 
Some of the aspects of the event have been already reported. \inlinecite{Papaioannou2022} and \inlinecite{Klein2022} studied the energetic particles and GLE produced by this event. The temperature and emission measure analysis of the observed wave was performed in the study of \inlinecite{Hou2022}.  
     
      \subsection{EUV Wave Kinematics and Dynamics}

\begin{figure}    
\centerline{\includegraphics[width=\textwidth]{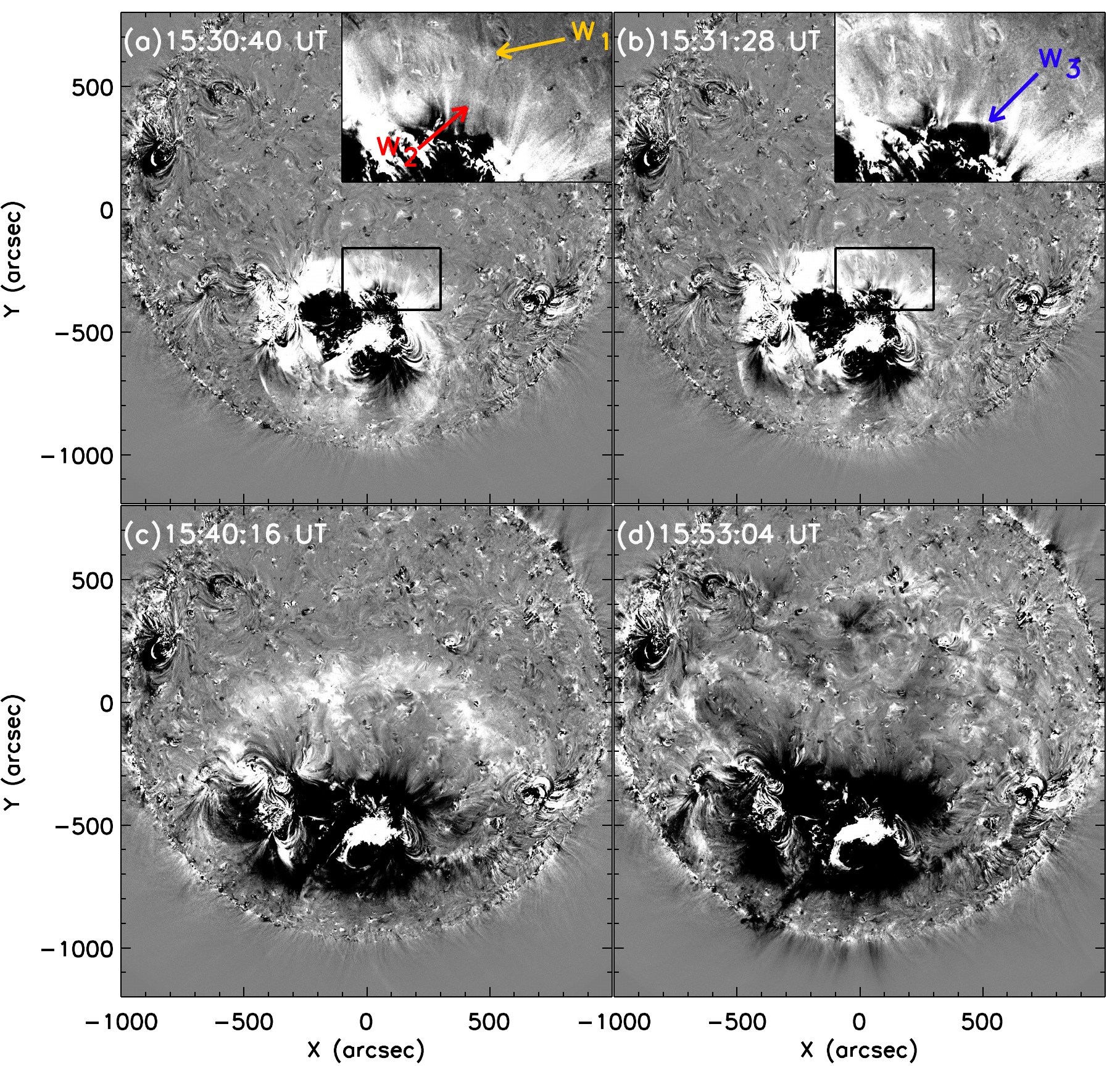}}
\caption{Evolution of the EUV wave in base difference images of AIA 193 \AA. The base image is chosen at 15:00:16 UT. The wave can be seen as the bright dome-shape structure. The {\it black box} in panels a and b is zoomed in the {\it upper right corner} of these images, where w$_1$, w$_2$, and w$_3$ are different wave-trains visible in this event. An animation of this figure is available as online supplementary material (\url{AIA193_28Oct2021_Fig1.mp4}).
}
\label{fig_wave_evolution193}
\end{figure}

On 28 October 2021, a giant EUV wave was
captured by the AIA telescope at its different EUV filters.
Because of the better visibility of the EUV wave in 193 \AA, we present the evolution of the EUV wave in this wavelength (shown in Figure \ref{fig_wave_evolution193}). The difference images are created by subtracting the pre-image at 15:00 UT when there was no activity in the active region. The appearance of the EUV wave is at approximately  15:25 UT in AIA 193 \AA. We find that initially the EUV wave is circular in shape. 
The southward propagation of the EUV wave is towards the limb while the northward propagation towards the disk center. Therefore, the EUV wave becomes more apparent near the disk center in the north direction as it evolves and less visible in the South due to perspective effects.
The EUV wave front is visible on the solar surface up to 15:50 UT and travel to the maximum distance of $\approx$ 1000$''$ from its origin site. Afterward, the wave becomes fainter. As the wave moved, we observe three wavefronts. Due to the strong strength of the wave towards the north, these fronts are clearly visible in that direction. These multiple wavefronts are labeled as w$_1$, w$_2$, and w$_3$ by yellow, red, and blue arrows, respectively, in the inset of Figure \ref{fig_wave_evolution193}a and b. Hereafter, we refer these wavefronts as ``wave trains'' as proposed in earlier observations \citep[for example,][]{Liu2012, Shen2022}.

\begin{figure}    
\centerline{\includegraphics[width=\textwidth]{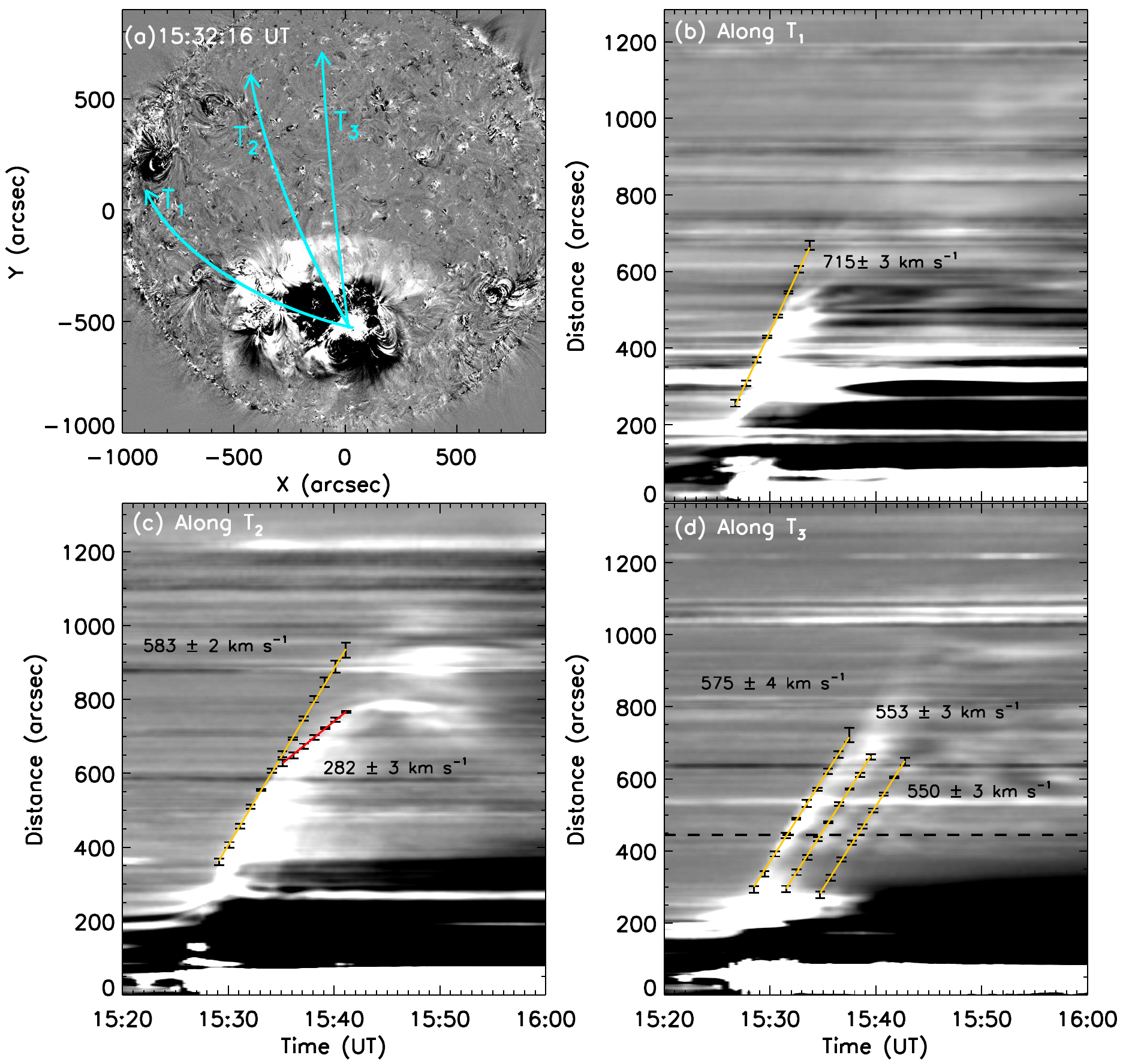}}
\caption{Base difference image (panel a) and time-distance plots in AIA 193 \AA~wavelength (panels b--d). Three slices T$_1$, T$_2$, and T$_3$ in different directions of the EUV wave propagation are shown in panel (a). The time-distance plots corresponding to these slices are displayed in panels b, c, and d, respectively.
The fitting for the velocity calculation of the fast mode wave and non-wave components are drawn by {\it yellow} and {\it red} lines, respectively. The intensity plot corresponding to {\it black dashed horizontal cut} is shown in Figure \ref{fig_wavetrain_wavelet}.}
\label{fig_wave_kinematics}
\end{figure}

Further, the kinematics of the observed EUV wave is analyzed. For this,  the time-distance analysis is done along three selected slices namely T$_1$, T$_2$, and T$_3$ towards the north direction, which are drawn in Figure \ref{fig_wave_kinematics}a. The results of these time-distance analyses are displayed in panels b, c, and d of Figure \ref{fig_wave_kinematics}. 
For the calculation of the speed, the leading edge of the wavefronts is identified visually. We compute the error associated with this identification by making three measurements at the same time and calculating the standard deviation. The error in the measurements is taken as three times the standard deviation and are shown in Figure \ref{fig_wave_kinematics}b, c, and d. Further, we fit the straight line through these selected data points and compute the average velocity of fast as well as slow (non-wave) components of the EUV wave. The reason behind the straight line fitting for the velocity computation is to highlight the fast and the slow (non-wave) components of the EUV wave.
In these directions, the  fast computed velocities are in wide spans and range from 550 to 715 \kms. In our chosen slices, the maximum velocity is in the T$_1$ direction ($\approx $ 715 $\pm$ 3 \kms). Along slice T$_2$, we observe two wavefronts, one has a velocity of about 583 $\pm$ 2 \kms\ and the other of about 282 $\pm$ 3 \kms.
According to the speed ranges suggested in the study of \cite{Chen2016}, we believe that the first wavefront is the fast-mode EUV wave and the 
second front is the non-wave component. The non-wave component can be explained by the magnetic field line stretching model proposed by \cite{Chen2002}. This non-wave component stops after reaching the distance of about 800$''$ at 15:42 UT. 
The non-wave component is visible only in the T$_2$ direction and not any other direction.
Inspection of slice T$_3$ reveals three parallel wave trains with velocities 575 $\pm$ 2 \kms, 553 $\pm$ 3 \kms, and 550 $\pm$ 3 \kms shown in panel d of Figure \ref{fig_wave_kinematics}. We find that as we move towards the outer edge of the EUV wave, the velocity is increasing and the outermost wave train has maximum velocity. From their speeds, we conclude that these multiple wavefronts are fast–mode EUV wave.

\begin{figure}    
\centerline{\includegraphics[width=\textwidth]{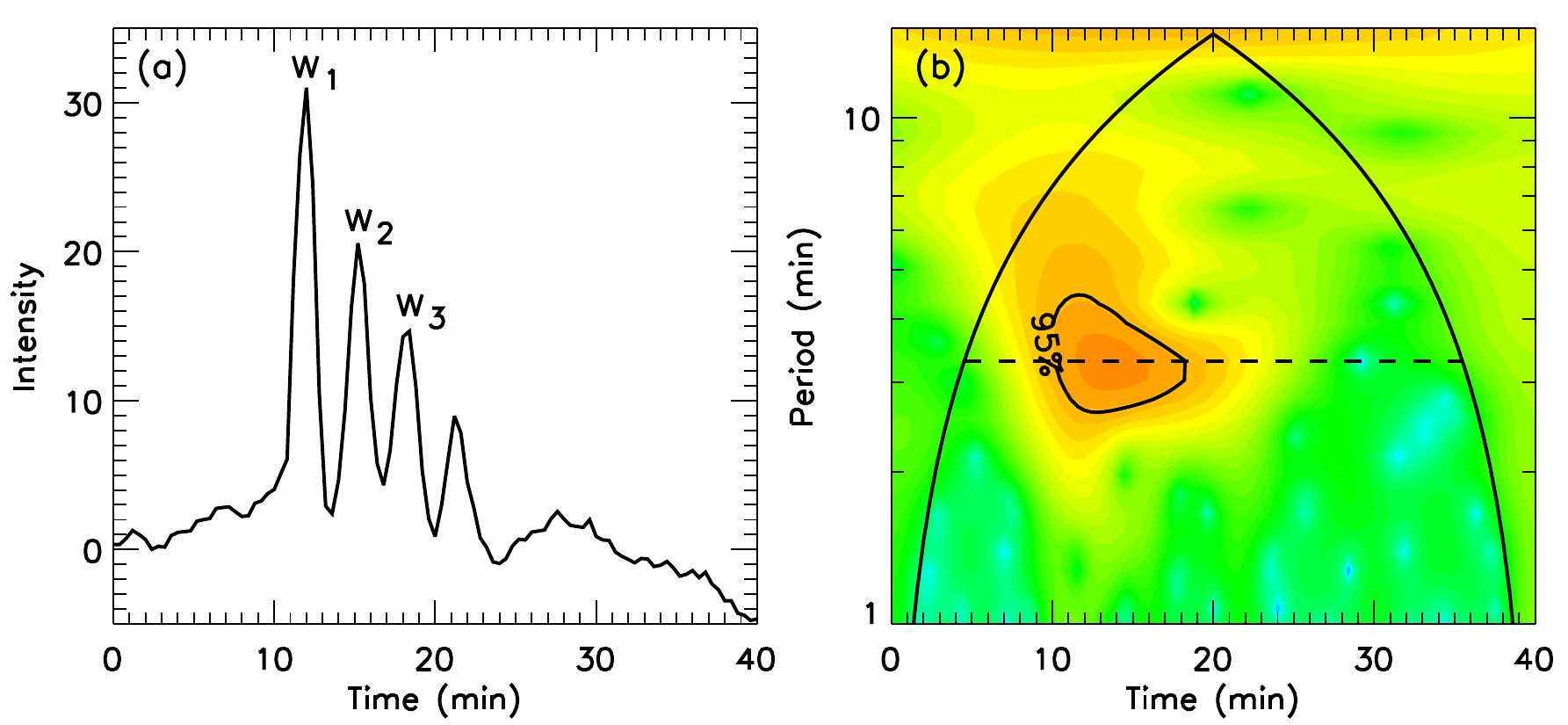}}
\caption{(a): Intensity plot along the {\it horizontal dashed line} in Figure \ref{fig_wave_kinematics}d. The w$_1$, w$_2$, and w$_3$ are the peaks corresponding to the wave-trains mentioned in Figure \ref{fig_wave_evolution193}. (b): wavelet corresponding to panel a showing a period of about 3.3 minutes. The period is labeled by a {\it horizontal dashed line}.}
\label{fig_wavetrain_wavelet}
\end{figure}

To explore the temporal evolution of the intensity of the wave trains, we explore  several cuts perpendicular to the time-distance diagram at different locations; the temporal evolution along the selected cut is shown in Figure \ref{fig_wavetrain_wavelet}aa.
This plot shows four peaks that correspond to four wave trains. However, in the difference images only three wave trains are promptly visible and the fourth wave train is very faint. We observe that the intensity of the wave trains are decreasing as we move along the time. Next, we check the periodicity of this intensity variation  using the wavelet analysis technique. The wavelet method for the period calculation is explained as follows. The wavelet transform returns wavelet scale information as a function of time, and it gives a smoothed version of the power spectrum. This method allows us to investigate the time dependence of periods within the observed data, i.e., the time localization of periods \citep{Torrence1998}. For the significance of time periods in the wavelet spectra, we take a significance test into account, and levels higher than or equal to 95\% are labeled as real. The significant region is bounded by the cone-of-influence (COI) region, which acts as an important
background for the edge effect for the given time range. The output of the  wavelet analysis is presented in Figure \ref{fig_wavetrain_wavelet}b. We find a period of 3.3 min with a confidence level of  95\%.

\subsection{Association of the EUV Wave with the CME}
Further in this section, we present the possible association between the EUV wave and the CME using the STEREO-A and LASCO observations.

\subsubsection{STEREO Observations}

\begin{figure}    
\centerline{\includegraphics[width=0.8\textwidth]{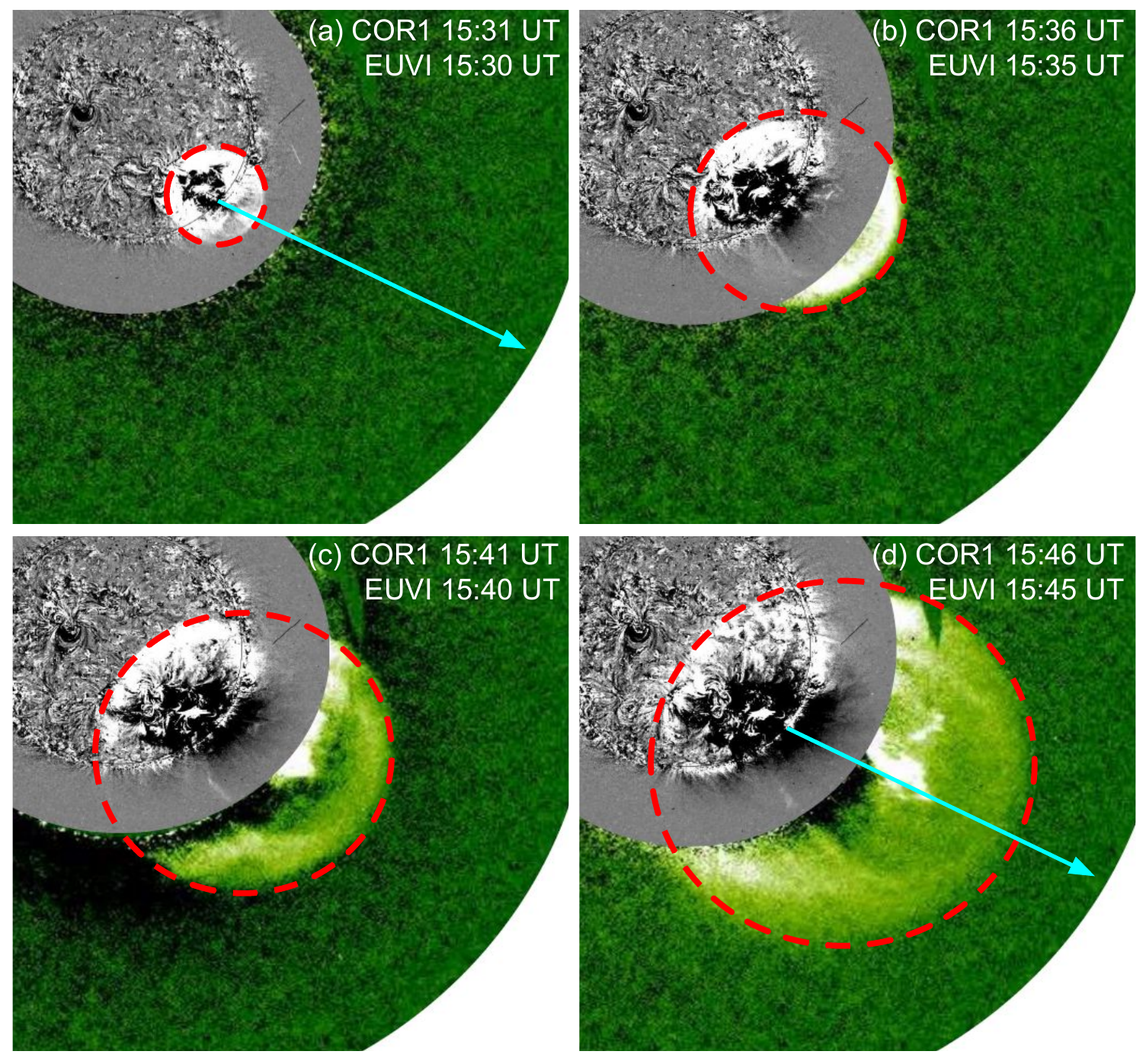}}
\caption{Evolution of the EUV wave observed with STEREO EUVI 195 \AA\ up to 1.7  R$_{sun}$ ({\it inner greyscale}) and the CME observed with COR1  for R $>$ 1.7 R$_{sun}$ ({\it outer green}). 
The time of each image is written on the {\it top right} of every panel. The {\it red circle} in all the images outlines the EUV wave front. 
The {\it cyan arrows} in panels a and d indicate the progression of the EUV wave in the corona ({\it inner greyscale}) 
as well as  the shock wave in front of the CME from 1.7 R$_{sun}$. These arrows  are used for the time distance diagram analysis given in Figure \ref{fig_timeslice_euvi_cor1}.
See the accompanying animation included as supplementary material (\url{STEREO_28Oct2021_Fig4.mp4}).}
\label{fig_image_euvi_cor1}
\end{figure}

The EUV wave was observed by the STEREO-A EUVI telescope in 195 \AA\ and the corresponding CME was registered by the COR1 coronagraph. The EUVI observes the Sun up to 1.7 R$_{sun}$ and COR1 observes the Sun after 1.5 R$_{sun}$. It means that between 1 R$_{sun}$ and 1.5 R$_{sun}$ we observe only the EUV corona and there is no white light data. This satellite combination provides us an excellent opportunity to compare the EUV wave with the CME, which is an important issue. For this purpose, similar to AIA data, we create the base difference images (base time 15:00 UT) for STEREO-A EUVI and COR1 data (base time 15:01 UT) and co-aligned these images. Four co-temporally over plotted images are displayed in Figure \ref{fig_image_euvi_cor1}. Further, we assume the EUV wave front has a circular shape as suggested by \cite{Gopalswamy2013} and fit a circle around its outer edge. This outer edge is drawn by a dashed circle in the same figure. Along with the assumption of circular evolution of the EUV wave and the spatial evolution of co-aligned and combined images, we infer that the EUV wave looks to move just ahead of the COR1 field-of-view (FOV) CME.

\begin{figure}    
\centerline{\includegraphics[width=\textwidth]{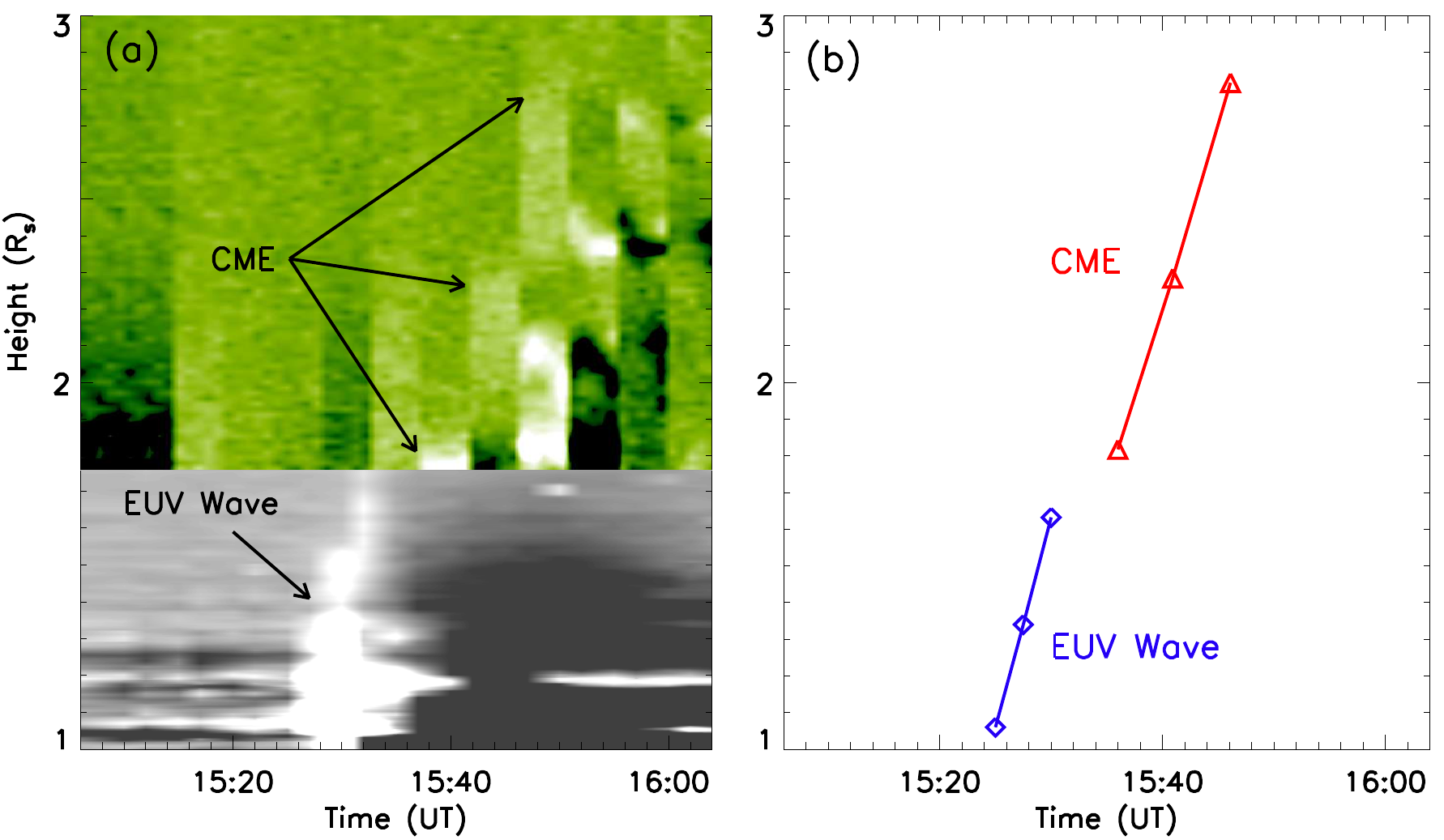}}
\caption{{\it Left panel}: The combined time-distance plot of the EUV wave ({\it greyscale}) and CME ({\it green color}) in STEREO EUVI 195 \AA~ and COR1 in the direction shown in Figure \ref{fig_image_euvi_cor1}a and d. The EUV wave and the CME are shown with {\it black arrows}. Base difference images of EUVI and COR1 are used for this analysis. The right panel shows the leading edges of the EUV wave ({\it blue}) and the CME ({\it red}) traced from the time-distance plot in panel a. 
}
\label{fig_timeslice_euvi_cor1}
\end{figure}

Next, the time-distance evaluation is performed for further validation of the EUV wave and CME relation. The slice selected for this purpose is drawn by the cyan arrow in Figure \ref{fig_image_euvi_cor1}a and d and the outcome of this is shown in Figure \ref{fig_timeslice_euvi_cor1}a. The the EUV wave visible in EUVI data and COR1 CME leading edge are labeled by arrows. Figure \ref{fig_timeslice_euvi_cor1}b displayed the measured leading edge points of the EUV wave and associated CME as a function of time until 3 R$_{sun}$. If the COR1 CME data points are extrapolated in the backward direction (or the EUV leading edge data points in the forward direction), we again find that the EUV leading front is ahead of the CME, which confirms the above finding drawn from Figure \ref{fig_image_euvi_cor1}.
Nevertheless, this conclusion is based on strong assumptions, the circular front of the CME in the disk plane is perhaps just observed by chance. 
However,  in the reconstructed CME using classical methods
\citep[i.e. the Graduated Cylindrical Shell model;][]{Thernisien2011} the front of the CME has a spherical shape like in the cone model.

\subsubsection{LASCO Observations}

\begin{figure}    
\centerline{\includegraphics[width=\textwidth]{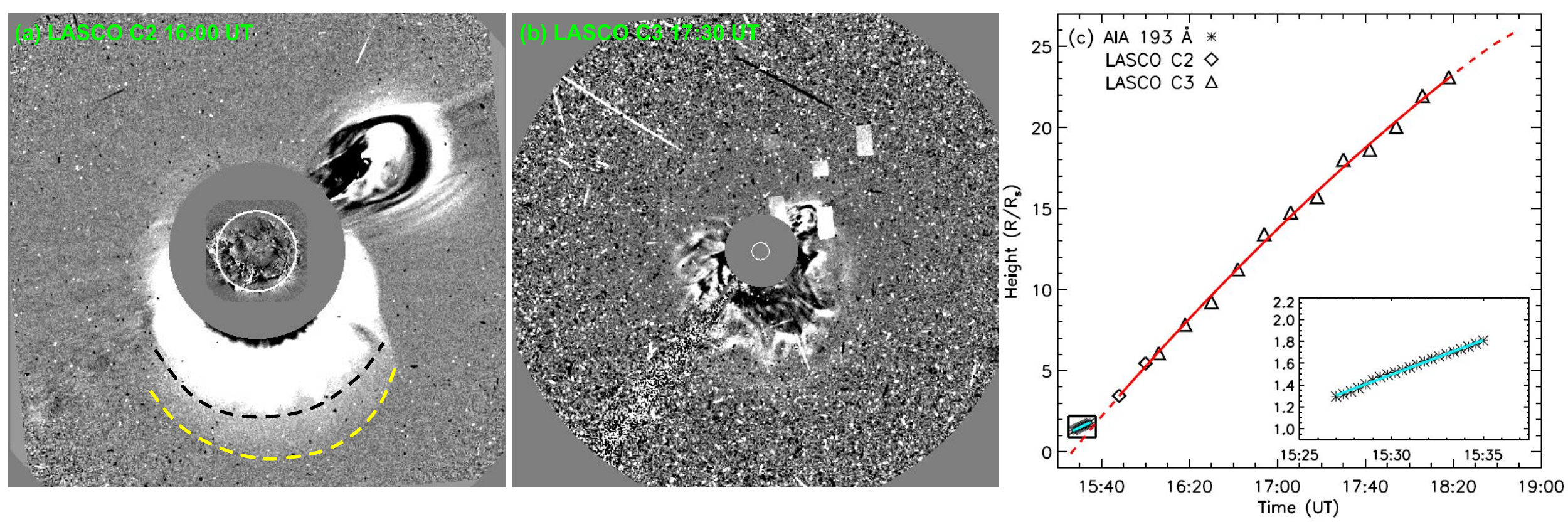}}
\caption{Difference images of LASCO C2 and C3 are presented in panels a and b, respectively. The CME leading edge and the shock are shown by {\it black} and {\it yellow dashed arcs}, respectively. 
Panel c displays the time-profile of the EUV wave with {\it asterisks} enclosed within the {\it black box}. The zoomed view of this {\it black box} is shown in the inset of the figure. These data points are the EUV wave fronts taken from AIA 193 \AA~ images until 1.4 R$_{sun}$ and fitted with linear function, shown by the {\it cyan line}. 
The {\it diamonds} and {\it triangles} are the data of the leading edge of the CME from LASCO C2 and C3, respectively. The {\it red solid line} is second order polynomial fit to the CME data and {\it red dashed line} is  extrapolation of this fitted curve. }
\label{fig_lasco_aia}
\end{figure}

An associated halo CME, originated from this AR, appeared in LASCO C2 FOV at 15:48 UT at a height of 3.45 R$_{sun}$. The leading edge of the CME remains in C2 FOV upto 16:24 UT. It  appeared in C3 FOV around 16:06 UT at a height of 6.08 R$_{sun}$ and was visible in its FOV for several hours. The snapshots of the CME in C2 and C3 FOV are shown in panels a and b of Figure \ref{fig_lasco_aia}. The C2 image reveals two structures: the diffuse one  followed by the  bright front. These two fronts are indicated by yellow and black arc drawings, respectively in Figure \ref{fig_lasco_aia}a. The yellow arc corresponds to the shock front and the black arc represents the CME leading edge. We have computed the height-time evolution of the CME propagation in C2 (diamond symbols) and C3 (triangle symbols) FOV and these measurements are plotted in panel c of the figure. The EUV wave observed by AIA 193 \AA\ temporal evolution is also overplotted in the same figure with asterisk symbols. Further, for the speed computation of the CME, we have fitted the second order polynomial with the data points, which is shown by a solid red line. The 
$\chi^2$ for this fitting is 1.14. From the fitting, the computed speed is 1240 \kms\ with an acceleration of -60 m s$^{-2}$. Next, we extend backward the second order polynomial fitting as shown by the red dashed line. The dashed line is slightly lower than the  EUV leading edge as observed in the case of lower height STEREO-A EUVI and COR1 instruments. Therefore, for this height-time comparison, we can guess that the EUV wavefront is always ahead of the CME leading edge.

\subsection{ Loop oscillations}

\subsubsection{Period Analysis}      
\begin{figure}    
\centerline{\includegraphics[width=0.9\textwidth]{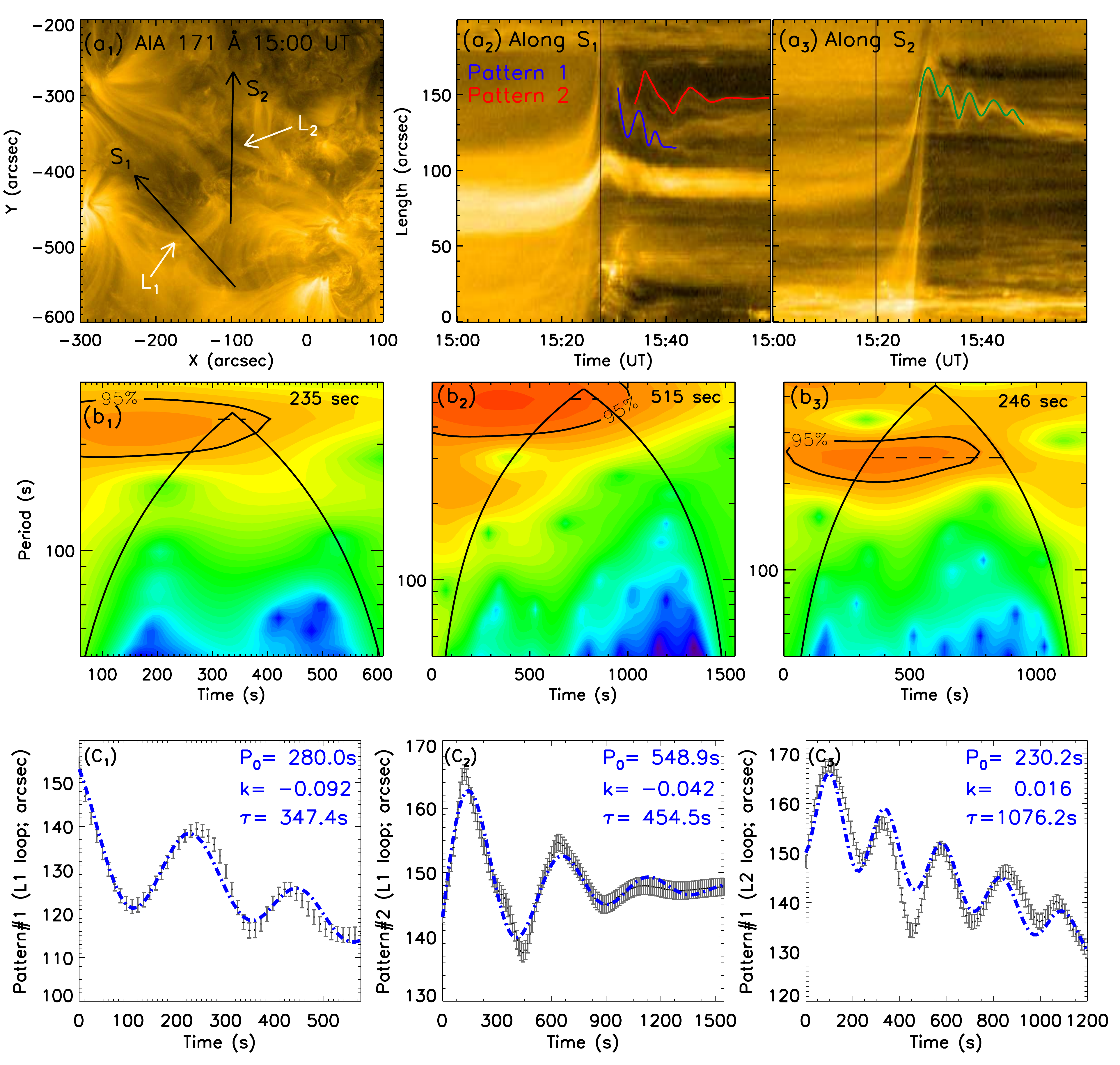}}
\caption{Slits S$_1$ and S$_2$ chosen for the time-distance analysis of loop systems L$_1$ and L$_2$, respectively, are shown in panel a$_1$ and corresponding time-distance plots in panels a$_2$ and a$_3$. The tracked patterns of the time-distance diagram along slice S$_1$, pattern 1 ({\it blue}) and pattern 2 ({\it red}) and S$_2$ ({\it green}) are over-plotted in panels a$_2$ and a$_3$. The corresponding wavelet analyses are shown in panels b$_1$, b$_2$, and b$_3$, respectively. The {\it horizontal dashed lines} are the corresponding periods. {\it Bottom panel}: Damped sinusoidal fitting of the oscillatory patterns corresponding to loop systems L$_1$ and L$_2$. The tracked oscillation patterns are shown in {\it black} and  
damped sinusoidal fits are over-plotted in {\it blue} (panels c$_1$ -- c$_3$). The fit parameters are annotated in the respective panels.
An animation of the oscillating loops corresponding to panel a$_1$ is included as supplementary material (\url{AIA171_28Oct2021_Fig7.mp4}).}
\label{fig_wavelet_s1_s2}
\end{figure}

As the EUV wave moves across the solar atmosphere, it interacts with the surrounding loops and originates the fast--mode kink oscillations (see the movies accompanied with Figure \ref{fig_wavelet_s1_s2}).
Due to the clarity of the oscillations, we have selected two loop systems, named L$_1$ and L$_2$, for further analysis. This analysis is performed using the time-distance plots of the loop oscillations. For the time-distance analysis, we have investigated several slits perpendicular to the loop systems and finally stayed two slits S$_1$ and S$_2$ perpendicular to the loop systems L$_1$ and L$_2$, respectively. These loop systems, slits, and their time-distance plots are shown in upper panel of Figure \ref{fig_wavelet_s1_s2}.
      
For the period calculations, we have tracked the oscillations patterns from the time-distance diagram. The identification of oscillating features has been purely done manually and shown in Figure \ref{fig_wavelet_s1_s2}a$_2$ and a$_3$. We find that there are two oscillating patterns along slit S$_1$ and one along slit S$_2$. The two patterns along slit S$_1$ are named as pattern 1 and pattern 2 and shown in blue and red colors, respectively, in panel a$_2$ of Figure \ref{fig_wavelet_s1_s2}. Similarly, the oscillation along slit S$_2$ is shown by green color in panel a$_3$ of the same figure. 
We have computed the period of these oscillations using wavelet analysis and fitting the damped sinusoidal function in the tracked patterns.
The results of the wavelet analysis are shown in Figure \ref{fig_wavelet_s1_s2}b$_1$, b$_2$, and b$_3$. The COI region is shown in black color.
From this technique, the period of oscillations for L$_1$ pattern 1 and pattern 2, and L$_2$ are found to be $\approx$ 235, 515, and 246 s, respectively. All the periods are marked by horizontal black dashed line in the middle panel of the figure. 

\begin{center}
\begin{table}
\begin{tabular*}{\textwidth}{cccc}
\hline
Parameters & \multicolumn{2}{c}{Loop System L$_1$} & Loop System L$_2$ \\
 & Pattern 1 & Pattern 2 & \\
\hline
$a_0$ (km) & 138.179 & 148.25 & 159.34 \\
$a_1$ (km~s$^{-1}$) & -0.038 & -0.0005 & -0.023 \\
P$_0$ (s) & 280 & 549 & 230 \\
k & -0.092 & -0.042 & 0.016 \\
$\tau$ (s) & 347 & 454 & 1076 \\
$\phi~(^{\circ})$ & 0.55 & 1.31 & 0.35 \\
Reduced $\chi^2*$ & 1.96 & 1.14 & 6.99 \\
\hline
\end{tabular*}
\caption{Table for different measured parameters of loop systems L$_1$ and L$_2$.}
\label{table}
* The reduced $\chi^2$ is calculated by $\chi^2$/DOF, where DOF is degree of freedom.
\end{table}
\end{center}

In addition to the wavelet method, we fit the identified oscillatory patterns with the damped sinusoidal function with the period varying with time according to the following equation suggested by  \inlinecite{Su2018}
\begin{equation}
    y(t)=a_0+a_1 \cdot t+a_2 \cdot e^{-t/\tau} \cdot cos\bigg(\frac{2\pi t}{P_0+kt}-\phi\bigg)
\end{equation}
Here $a_0$, $a_1$, and $a_2$ are constants related to the amplitude of the oscillation, $\tau$ is damping time, $\phi$ is the initial phase, and $P_0+kt$ is the empirical expression for the oscillation period where the constant `$k$' represents a linear variation of the period in time. For fitting the damped sinusoidal function with the observed pattern, we use the mpfitfun.pro procedure that employs the Levenberg-Marquardt least-squares fitting to the user defined functions. The function is available in the solar software. We find the period P$_0$ to be varying in the range of $\approx$ 230--549 s, which is in agreement with that obtained by the wavelet analysis. The fitting parameters for different loop systems are given in Table \ref{table}. 

It is interesting to note that the shoulders (the part of the oscillation which is not well fitted with sinusoidal fit) are visible and show a departure from the fit in C$_2$ as well as C$_3$. 
This may be due to the mixing of several indistinguishable adjacent oscillating bright loops.
That is the reason to call them a “loop system”. Due to different 
magnetic field values corresponding to different loops, we do not expect the same period, or amplitude. Another explanation can be that the slits may not be completely perpendicular to the loop systems, although we have computed such oscillatory features for different slit inclinations and used the time-slices that show the best oscillatory signatures.

\subsubsection{Temperature, Density and Magnetic Field}

\begin{figure}    
\vspace{-2.5cm}
\centerline{\includegraphics[width=\textwidth]{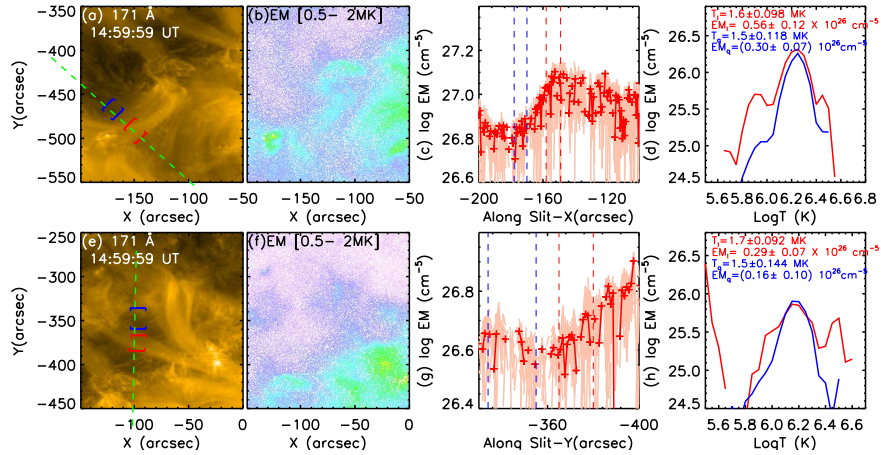}}
\caption{Thermal characteristics of the oscillating loops L$_1$ and L$_2$ in the top and bottom panels, respectively. (a) A segment of the AR containing loop L$_1$ in 171 {\AA} and the corresponding EM map in 0.5--2.0 MK shown in panel b. (c): EM values along the {\it dotted slant line} (drawn in panel a), with the {\it vertical dotted lines} in {\it red (blue)} marking the region within (outside) the investigated loop (also marked in panel a). (d): Averaged EM[$T$] distribution for the region identified as inside ({\it red}) and outside ({\it blue}) of the loop. Similarly bottom panel shows the thermal characteristics for loop L$_2$.}
\label{fig_T_EM_L1_L2}
\end{figure}

We have investigated the thermal characteristics of the plasma contained in the coronal loops L$_1$ and L$_2$ as presented in Figure~\ref{fig_T_EM_L1_L2}. In this regard, the emission measure (EM) maps have been synthesized using the modified sparse differential emission measure (DEM) inversion technique \citep{Cheung2015, SuY2018} which allows to effectively constrain the EM values employing the AIA/SDO observations in six wavelength channels namely 94, 131, 171, 193, 211, and 335 {\AA}. We have synthesized the EM maps in the temperature range of 0.5 -- 30 MK with a temperature interval of $\Delta$log$T$=0.05. However, the temperature range of 0.5 -- 2MK can adequately represent the thermal characteristics of the plasma contained in the coronal loops away from the flaring region as shown in Figure~\ref{fig_T_EM_L1_L2}b and f. 
 The DEM inversion method by \citet{SuY2018} further allows deriving the  uncertainty in the estimated EM values by employing a Monte-Carlo simulation. Essentially, the method estimates the EM distribution for the intensity values which are slightly modified by adding random noises for a user-defined number of times (n=100 in this work). Therefore, for every pixel of the image, we derived 100 additional EM[$T$] distributions in addition to that estimated from the actual observed intensities.
The EM profiles determined over a line (dotted lines in Figure~\ref{fig_T_EM_L1_L2}a and e) spanned across the loop structures reveal that the EM values inside the loops are approximately two-times higher than those determined in the close vicinity outside the loop e.g. for L$_1$, the EM values inside and outside the loops are determined to be (log EM) 26.8 and 27.1. 
 Such a trend can also be noticed in all the EM profiles derived from the additional EM maps, over-plotted in light red.
Next, the EM[$T$] distribution corresponding to the plasma inside and outside the loop has been estimated by averaging the EM[$T$] values in respective temperature bins at all the pixels lying within the visually identified boundaries (red and blue vertical dotted lines in Figure~\ref{fig_T_EM_L1_L2}c and g). From the EM[$T$] distribution, we further determine the representative EM-weighted temperature ($<T_{EM}>$) in the temperature range log$T$ (k)=[5.8, 6.7] using the following equation.
\begin{equation}
    T_{EM}=\frac{\sum_jT_j\times {\it EM}[T_j]}{\sum_j EM[T_j]}
\end{equation}
Further, the EM value inside and outside the coronal loop is determined by taking an average of the EM values within the aforementioned temperature range. In this way, the $<T_{EM}>$ and  $<EM>$ values within (and outside) the loop system L$_1$ is derived to be 1.6 $\pm$ 0.098 MK (1.5 $\pm$ 0.118 MK), and 0.56 $\pm$ 0.12 (0.3 $\pm$ 0.07) $\times$ 10$^{26}$ cm$^{-5}$. 
 We have also determined the  $<T_{EM}>$ and  $<EM>$ values for the aforementioned positions and temperature bins in all the additional EM maps derived using the Monte-Carlo technique. The standard deviation of the estimated values is assigned as the uncertainty in each parameter.
Similarly, the temperature and EM values inside (outside) the loop system L$_2$ are estimated to be 1.7 $\pm$ 0.092 MK (1.5 $\pm$ 0.144 MK), and 0.29 $\pm$ 0.07 (0.16 $\pm$ 0.10) $\times$ 10$^{26}$ cm$^{-5}$. 

Now, the densities inside and outside the coronal loops are calculated by the following equation:
\begin{equation}
n=\sqrt{\frac{EM}{s}},
\label{eq3}
\end{equation}

where EM, $n$, and $s$ are the emission measure, density, and  depth of the coronal loop, respectively. By putting the computed values of emission measure inside (EM$_{in}$) and outside (EM$_{ex}$) the loop systems the densities $n_{in}$ and $n_{ex}$ can be calculated. Now, for loop system L$_1$, the derived values of $n_{in}$ and $n_{ex}$ are 2.101 $\times$ 10$^8$ cm$^{-3}$ and 1.797 $\times$ 10$^8$ cm$^{-3}$, respectively. Similarly, for loop system L$_2$, these values are 1.630 $\times$ 10$^8$ cm$^{-3}$ and 1.025 $\times$ 10$^8$ cm$^{-3}$, respectively.
Using these density values, the ratio of densities inside and outside L$_1$ and L$_2$ are found to be 1.17 and 1.59, respectively.
Since the uncertainties are $\approx$ 25$\%$ (3$\%$) for the EM inside (outside), the error in densities can be expected to be at a maximum of $\approx$ 5$\%$. The density estimated inside is $\approx$ 1.6 times higher than outside. 

\begin{figure}    
\centerline{\includegraphics[width=\textwidth]{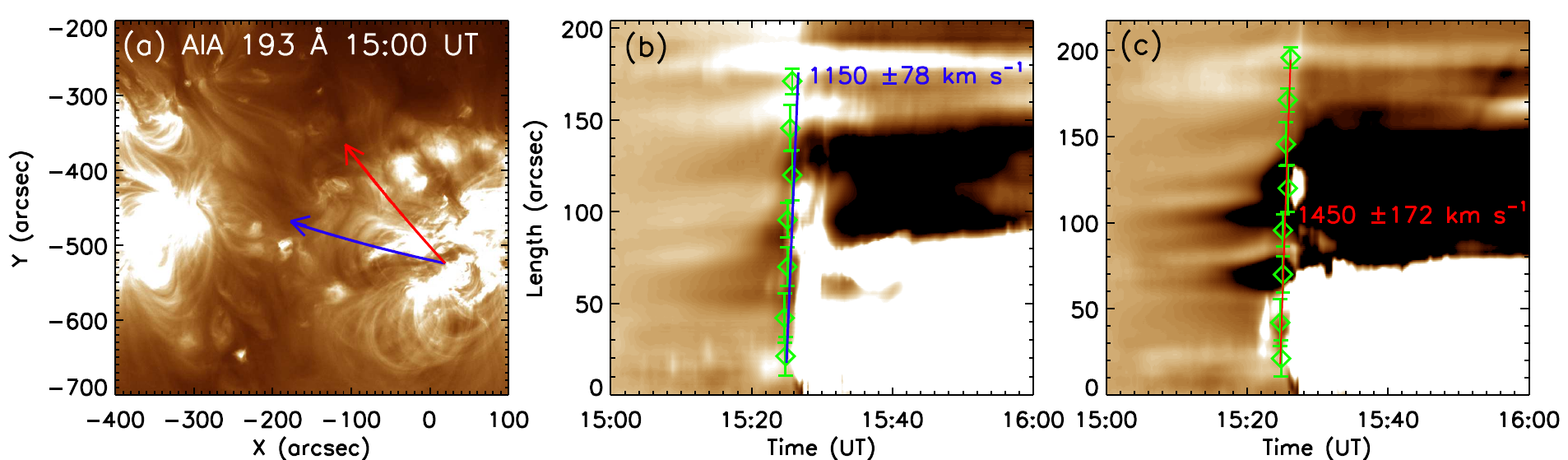}}
\caption{Slices chosen to find the speed of the fast magneto-acoustic wave towards loop systems L$_1$ and L$_2$ shown by {\it blue} and {\it red} colors, respectively. The corresponding time-distance diagrams are shown in panels b and c, respectively. The {\it green symbols} in panels b and c are the errors in the identification of the leading edge of the wave which is taken as three times the standard deviation from the three repeated measurements at the same time. The {\it blue} and {\it red lines} are the fitting in the identified data points.}
\label{fig_fast_speed}
\end{figure}

Further, we have computed the ratio of the densities inside ($n_{in}$) and outside ($n_{ex}$) the EUV loops theoretically using the following formula suggested by  \inlinecite{Aschwanden2011}. 
\begin{equation}
\frac{n_{in}}{n_{ex}}= \frac{1}{2}\bigg(\frac{L_{exc}}{T_{exc}}\frac{P_{kink}}{L_{osc}}\bigg)^2-1
=\frac{1}{2}\bigg(v_{exc}\frac{P_{kink}}{L_{osc}}\bigg)^2-1, 
\label{eq4}
 \end{equation}

where P$_{kink}$ and L$_{osc}$ are the period and length of the coronal loop, respectively. L$_{exc}$ and T$_{exc}$ are the distance of the loop apex from the flare location and the time at which the EUV wave reaches up the loop apex. Dividing both these variables, we get the propagation velocity of the global fast magneto-acoustic wave when it reaches the loop, which is given by $v_{exc}$. To derive this velocity, we choose two slices in the direction of loop system L$_1$ and L$_2$, which are shown in Figure \ref{fig_fast_speed}a with red and blue colors, respectively. The time-distance diagram corresponding to these slices are displayed in panels b and c, respectively. From the time-distance diagram, we track the fast magneto-acoustic wave and fit a linear function $L(t)=v_{exc}t+L_0$, where $L_0$ is a constant. From this fitting, we get that the velocity $v_{exc}$ towards loop systems L$_1$ and L$_2$ are $\approx$ 1150 $\pm$ 78 and 1450 $\pm$ 172 km s$^{-1}$, respectively. 
Now, for the period, we use two different methods as explained above. Let us first calculate the density ratio using the P$_{kink}$ values from the wavelet method. 
We want to remark that the length of the loops whose oscillations are tracked as pattern 2 is ambiguous. While measuring this length, the footpoints of these loops are not very clear, which does not allow us to calculate their correct length. Therefore, we leave pattern 2 for further calculations. The measured lengths (L$_{osc}$) of loop systems L$_1$ and L$_2$ are $\approx$ 115 and 175 Mm, respectively. Inserting the values of P$_{kink}$ from the wavelet method and L$_{osc}$ in Equation \ref{eq4}, we get that the density ratios for loop system L$_1$ and L$_2$ are 1.76 and 1.08, respectively.
Now, we find the density ratios using the P$_{kink}$ values from the fitting method. From these measurements, the derived density ratios are 2.92 and 0.82 for loop systems L$_1$ and L$_2$, respectively. The density ratio 0.82 indicates that the density outside (n$_{ex}$) the loop is larger than inside (n$_{in}$) the loop, which is not physically possible.
Here, we would like to mention that there is an uncertainty in P$_{kink}$, v$_{exc}$, and L$_{osc}$ computation, which could be responsible for the error in the density ratio estimation. 

The coronal magnetic field strength ($B$) is estimated using the formula suggested by \inlinecite{Roberts1984} and \inlinecite{Aschwanden2011}, which is given as follows:
\begin{equation}
    B=\frac{L_{osc}}{P_{kink}}\sqrt{8\pi\mu m_p n_{in}\Big(1+\frac{n_{ex}}{n_{in}}\Big)} ,
\end{equation}
where $\mu$ is the average molecular weight for coronal abundances which is equal to 1.2 \citep{Verwichte2013, Guo2015} and $m_p$ = 1.67$\times$10$^{-24}$ g is the mass of the proton.
For this calculation, the densities $n_{in}$ and $n_{ex}$ are taken from the DEM method. Again, we can estimate different values for the magnetic field strength using the P$_{kink}$ values from the wavelet and the fitting method. If we put P$_{kink}$ from the wavelet analysis, the values of $B$ are 6.85 and 8.22 G for L$_1$ and L$_2$, respectively. Similarly, if we take P$_{kink}$ from the fitting method, these values are found to be 5.75 and 8.79 G. It can be noted here that the values of the magnetic field strength are consistent with both the methods. Therefore, we can say that the magnetic field strength ranges from 5.75 -- 6.85 G for loop system L$_1$ and 8.22 -- 8.79 G for loop system L$_2$.

It may be noted that there is a difference between the ratio of the densities estimated inside and outside the loop systems as obtained from the two methods (1.17 and 1.59 from one method (Equation \ref{eq3}), while 1.76 and 1.08 from the other method (Equation \ref{eq4})). The reason for such a difference may be that the estimations from the first method, which employs the EM maps, depends on the selection of coordinates for the loop interior and exterior (see Figure \ref{fig_T_EM_L1_L2}). On the other hand, although the second method (Equation \ref{eq4}) does not require the selection of the position on the loop, the identification of oscillatory patterns is subject to uncertainties. Therefore, we have provided a range of densities, as well as subsequently estimated the magnetic field values for the loop systems. 

\subsection{Filament Eruption}
\label{sect_filament}
\begin{figure}    
\centerline{\includegraphics[width=\textwidth]{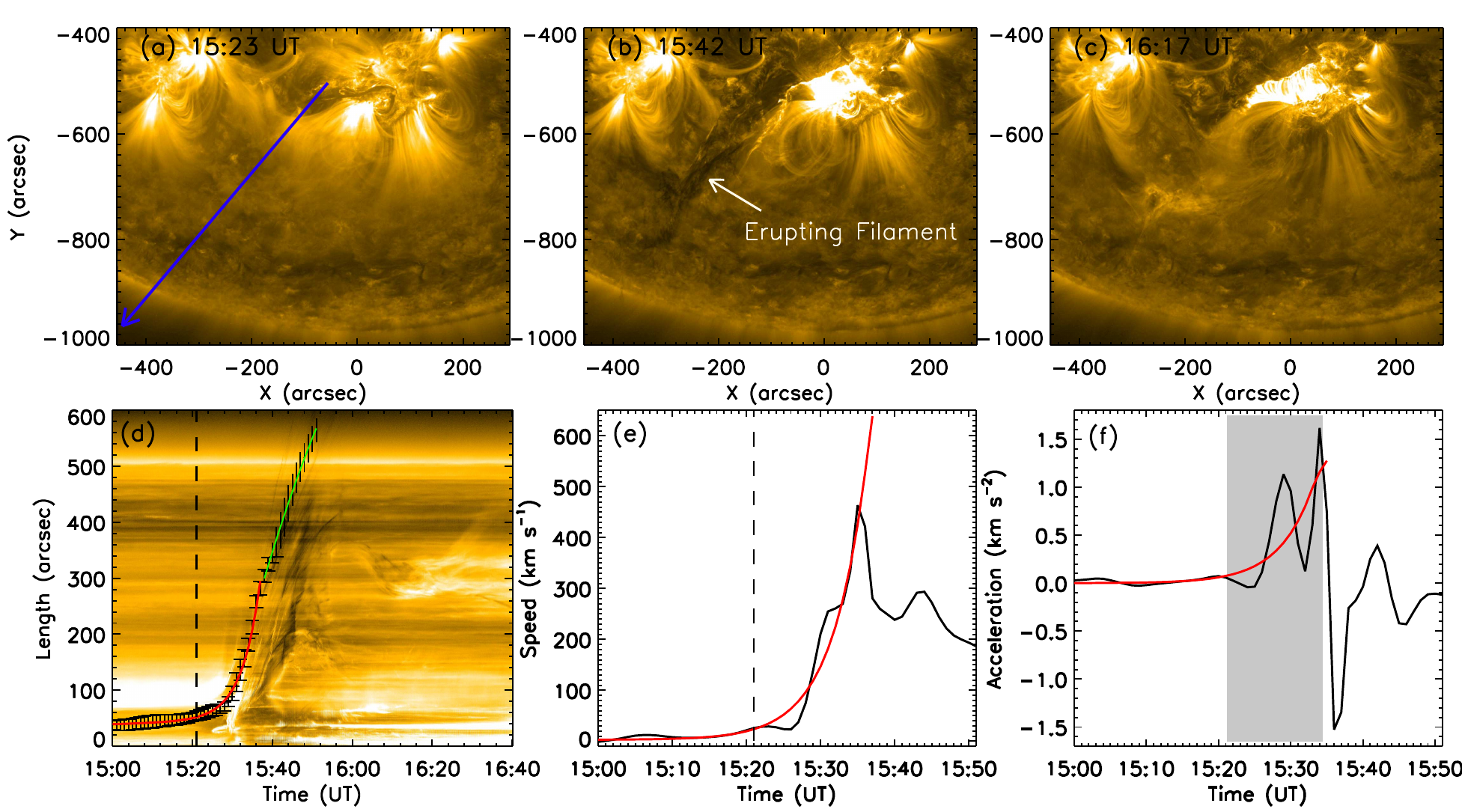}}
\caption{Evolution of the filament eruption in AIA 171 \AA~(upper panel). Panel d presents the time-distance diagram in the direction shown by the {\it blue arrow} in panel a. 
The {\it red curve} is the fitting of these data points with a combination of a linear and exponential function upto the acceleration phase and the {\it green curve} is the fitting with a second order polynomial after the acceleration phase (i.e. deceleration phase) of the eruption. The error bars in the filament height are drawn in panel d. Panels e and f display the speed and acceleration, respectively, with the tracked data points ({\it black}) and the fitting ({\it red}). A {\it dashed vertical line} in panels d and e is the start of acceleration phase of the eruption. The acceleration phase of the filament is highlighted by grey in panel f. An animation with same FOV accompanies this figure (\url{AIA171_28Oct2021_Fig10.mp4}).}
\label{fig_filament_kinematics}
\end{figure}

 A long (about $\approx$ 170 Mm) and relatively thin ($\approx$ 8 Mm) filament was situated in the active region along the polarity inversion line, which erupted on 28 October 2021 at 15:21 UT. The evolution of the eruption in AIA 171 \AA\ is shown in Figure \ref{fig_filament_kinematics}a -- c. In panel b of the figure, the erupting filament is indicated  by an arrow. As a result of this eruption, we observe two parallel flare ribbons, which separated as a function of time  as expected in the standard CSHKP model \citep{Carmichael1964, Sturrock1966, Hirayama1974, Kopp1976} (please see the accompanying movie \url{AIA171_28Oct2021_Fig10.mp4}).

The time-distance plot of the erupting filament is displayed in Figure \ref{fig_filament_kinematics}d. The direction of the selected slice is drawn in panel a of the same figure. This time-distance diagram is used to derive the height of the leading edge of the filament. 
After tracking the erupting filament, we apply the Savitzky-Golay filter to smooth the tracked data as suggested by \cite{Byrne2013}. The errors in the leading edge identification are estimated by making three measurements at the same times and calculating the standard deviation. Then, the error in the measurements are taken as three times the standard deviation.
Further, using the data of this height-time plot, we compute the speed and acceleration of the eruption. The combination of the linear and exponential functions of the form $ae^{bt}+ct+d$ is fitted with the selected height-time data sets. The a, b, c, and d are the constant parameters to be found by the fitting. The tracked data points and the fitting curve are shown by the black diamonds and red line, respectively, in Figure \ref{fig_filament_kinematics}d. Based on this fitting, we compute the onset time of the eruption using the formula 
$t_{start}=\frac{1}{b} ln (\frac{v_t}{ab})$, where $v_t=c\approx 6$ km s$^{-1}$, the linear speed of the filament eruption \citep{Cheng2020, Chandra2021, Devi2021}. From this equation, the
approximate onset time is 15:21 UT, which is shown by the vertical dashed line in panels d and e of the same figure. 

Next, we compute the speed and acceleration of the eruption. 
The velocity and the acceleration in the plane of the sky are obtained from the 
derivative of the smoothed distance with time. This velocity and acceleration are shown in Figure \ref{fig_filament_kinematics}e and f.
In these plots, the black and red curves correspond to the speed and acceleration derived from the manually tracked data and  fitting, respectively.
The acceleration phase of the eruption 
stays until approximately 15:34 UT and is highlighted by the grey shaded region. After about 15:34 UT, the filament is seen to decelerate.
The average (maximum) speed and acceleration of the erupting filament are 195 (640) \kms~and 0.5 (1.3) km s$^{-2}$,  respectively.
In the deceleration phase of the filament eruption, we fit a second order polynomial. This fit is shown with a green line in panel d of the figure. The deceleration of the eruption is 0.10 km s$^{-2}$.

\begin{figure}    
\centerline{\includegraphics[width=\textwidth]{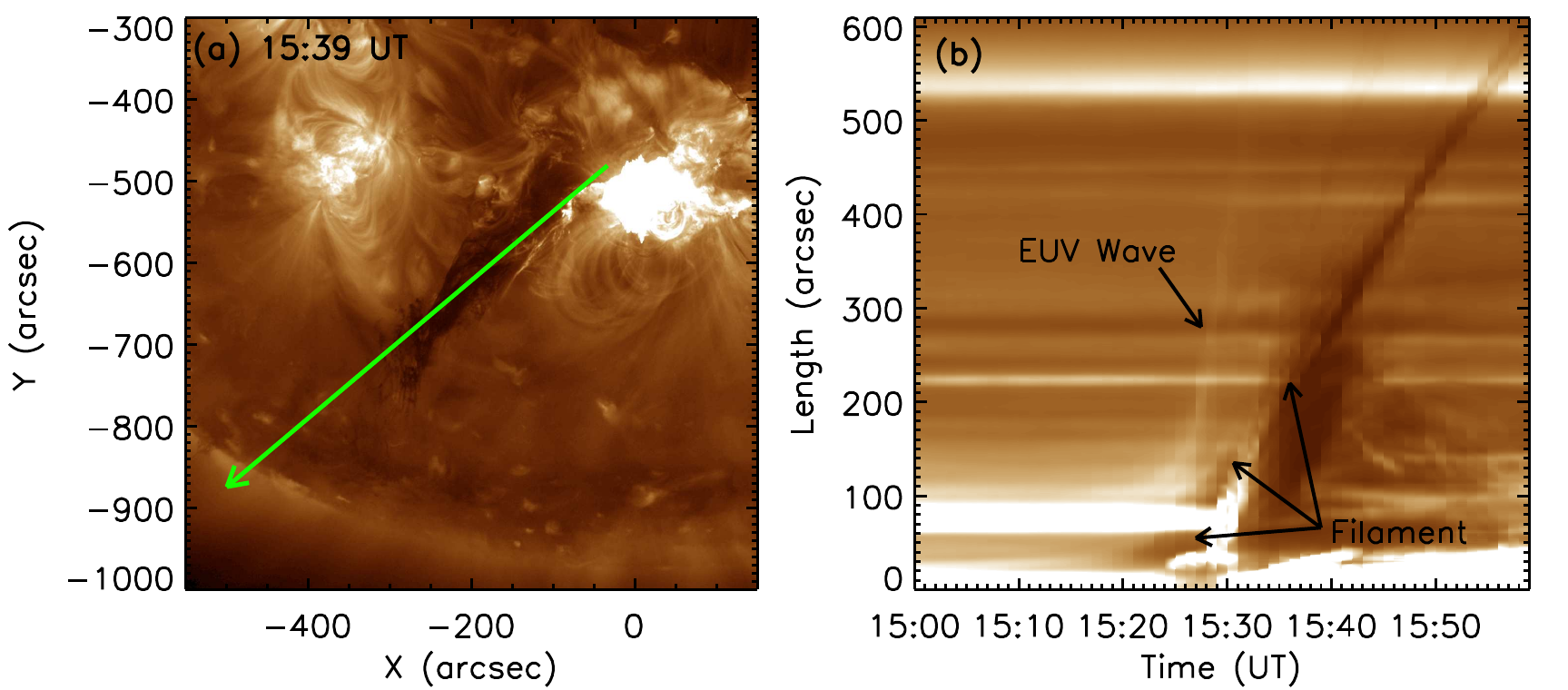}}
\caption{Image of AIA 193 \AA~ showing the slice using a {\it green arrow} in the direction of the filament eruption. The time-distance diagram is presented in panel b which shows the EUV wave and the filament using {\it black arrows}.}
\label{fig_timeslice_wave_filament}
\end{figure}

Figures \ref{fig_timeslice_wave_filament}a and b demonstrate the connection between the filament eruption and EUV wave. The data used here 
is AIA 193 \AA. In the time-distance plot, the EUV wave and the erupting  filament are marked by black arrows. This figure shows that the EUV wave and the filament eruption started closely in time. Moreover, as the eruption evolves, the filament eruption and the EUV wave separate. 
The EUV wave becomes faint at approximately 
15:30 UT after travelling a distance of 400$''$ from its origin.

\section{Discussion} 
      \label{S-discussion} 
  
 The origin and the nature of EUV waves were initially proposed
 using EIT data on board Solar and Heliospheric Observatory (SOHO)
 about two and half decades ago \citep{Moses1997, Thompson1998}. Later on, the multi-viewpoint observations of these phenomena were examined by instruments on board STEREO A and B satellites \citep{Patsourakos2009, Patsourakos2012, Delannee2014, Chandra2021}. However, due to the poor spatial and temporal resolution of the  above instruments, 
 we can not distinguish the several fine structural details of EUV waves. After the launch of the SDO satellite, the AIA telescope provides better spatial and temporal resolution data, which  allows us to investigate the nature of EUV waves in more detail. In this study, we analyze the two view-angle data sets available from the AIA, and STEREO-A EUVI and COR1 instruments.   
 
 The two observed  components of the EUV wave,
 the fast mode wave and non-wave components are explained by the hybrid model proposed by \citet{Chen2002}. According to their model, the fast mode component is the counterpart of the coronal Moreton wave and the slower non-wave component is the result of the  stretching of magnetic field lines due to the existing filament eruption.
 We find that the speed of the fast-mode component ranges from 575 to 715 \kms\ and the speed of the non-wave component clearly observed in one direction is 282 \kms. 
 The speed of the non-wave component depends on the overlaying magnetic field configuration.
 If the overlying magnetic field structure is majorly stretched in the horizontal direction, the difference between the speed of the fast mode EUV wave and the non-wave component would be lower. This case is similar to the case reported in the study of \citet{Chandra2021}. 
  
On the origin of the EUV wave, we explore the possibility of its association with flares, filament eruptions, and CMEs. Some studies suggest that the EUV waves are originated by a  pressure pulse generated during the impulsive or peak phase of the flare. In our case, we find that the flare peak time from GOES X-ray is around 15:35 UT (in hard X-ray the peak time is approximately
15:28 UT, \citep[see][]{Klein2022} and the filament eruption onset is at 15:21 UT. If we compute the EUV wave location at the flare peak time, we see that it was at a height of approximately
(around 300$''$ at hard X-ray peak time) from the flare origin.
 Furthermore, we compare the spatial location of the EUV wave with the CME observed by the STEREO-A spacecraft by its
 EUVI 195 \AA\ and COR1 instruments and find that the EUV wave is just above the CME. This possibility is further confirmed by the comparison of the EUV wave height with CME height
 evolution both in STEREO and LASCO data sets. This suggests that the EUV wave of 28 October 2021 is driven by the filament eruption or CME instead of a  flare pressure pulse. Our  results support the earlier findings  of \citet{Chen2006, Kwon2014,Kwon2017}.
 \citet{Chen2006} did a statistical study of 46 flare events and concluded that the EUV waves appear only when  flares are accompanied by  CMEs. \citet{Kwon2014,Kwon2017} studied the outer envelop of halo CMEs using  STEREO--A, STEREO--B, and SDO.
 and concluded that the outer envelop of a CME corresponds to a fast--mode EUV wave. Furthermore, they explain their observation using geometric models.
 
For the interpretation of the observed wave trains of this EUV wave, we explore the following possibilities suggested in the literature:  the EUV waves are (i) due to the quasi-periodic pulsations (QPP) observed in flares \citep{Liu2012, Shen2013, Ofman2018, Zhou2022}, (ii) due to the oscillations in surrounding magnetic field stretched by an erupting flux rope \citep{Shen2022}, and (iii) due to the untwisting motion of the erupting filament \citep{Miao2019, Shen2019}. In our case, since the flare peak or impulsive phase does not match with the time of the appearance of the EUV wave,
the possibility of the QPP for the wave train creations cannot be considered as their origin. We very clearly observe
the oscillations in the nearby loop systems, which suggest the possibility that the EUV wave trains are due to oscillations in the surrounding magnetic field. The third possibility of the wave trains may be due to the untwisting of the erupting filament as suggested in the study of \citet{Shen2019}. However, in our case, we do not observe clear evidence of  this  untwisting.
Therefore, we cannot confirm this possibility. In addition to the above, \citet{Chen2002} observed multiple wavefronts of an EUV wave in their numerical simulations, and  
our results agree with their simulations.  
 
In the present event, the observations of the fast-mode EUV wave components provide us an opportunity for the computation of the physical parameters of oscillating loops. The fast-mode wave moved in all directions and triggered the MHD oscillations in the nearby loop systems particularly in the north direction. 
Here, we focus only on the loop systems L$_1$ and L$_2$, where the oscillations are clearly visible. Using  DEM analysis, we estimate the electron density inside and outside, and the temperature of the selected loop systems. The values of the electron densities inside and outside are 2.101 $\times$10$^8$ and 1.797 $\times$10$^8$ cm$^{-3}$, and 1.630 $\times$10$^8$ and 1.025 $\times$10$^8$  cm$^{-3}$; while the temperatures are 1.6 MK and 1.7 MK for the L$_1$ and L$_2$ loop systems, respectively.
  
Furthermore, using  magnetoseismology, we have also computed the density ratio (the ratio of density inside and outside the loop). The computed density ratios for the loop systems L$_1$ and L$_2$ are 1.76 and 1.08 respectively. The density ratios obtained by the emission measure technique are 1.17 and 1.59 for  the L$_1$ and L$_2$ loop systems, which are consistent with values derived by magnetoseismology. 
  Here, we would like to mention some of the results of density ratios obtained with the magnetoseismology procedure: \citet{Guo2015} studied the oscillations of the loops triggered by the EUV wave during the event of 11 April 2013 and computed the density ratio of the oscillating loop. The calculated value was 1.3. Another study done by \citet{Su2018} reported the ratio  to be between 4 to 7. Apart from the EUV wave triggered oscillations, studies are done on  flare initiated loop oscillations and computed  density ratio in loops \citep[e.g.][]{Nakariakov99, Aschwanden2011, Srivastava2013} with values ranging  
  from 1--12. The difference in density ratios may be due to the fact that  the magnetoseismology results depend upon the period, length of the oscillating loops, and the speed of EUV fast magnetoacoustic wave.

  Another useful application of magnetoseismology is to estimate the magnetic field strength of the oscillating loops or
  filaments. Our values range from 5.75 to 8.79 G. In the studies of loop oscillations, the estimated magnetic field strength varies from 4 to 43 G \citep{Nakariakov2001, Aschwanden2011, Guo2015, Su2018, Zhang2022}. If 
  the derived magnetic field values using the filament oscillations is considered, the range of the computed coronal magnetic field is 3--30 G \citep{Hyder1966, Isobe2006, Gilbert2008, Gosain2012, Luna2014, Arregui2018, Devi2022}. In summary, our estimated values are  consistent with earlier findings.

 The study of the kinematics of the filament eruption can shed a light on the  
 mechanism  triggering its eruption \citep{Demoulin2010, Schmieder2012}.
  Due to this we have tried several fitting functions with the erupting filament data (the details about this fitting are discussed in Section \ref{sect_filament}). For this analysis, we find the filament starts to rise with a low v $\approx$ 6 \kms) and approximately linear speed  approximately upto around 15:21 UT. After that, the filament exhibits an exponential rise with an average speed of approximately
  195 \kms~and 
  an acceleration which increases non-linearly. This exponential rise of the filament indicates the presence of a torus instability. This type of eruption behavior is found by \inlinecite{Cheng2020} in the sample of twelve events and also in case studies  by \citet{Schrijver2008,  Devi2021, Chandra2021}. 
   
\section{Conclusion}

      \label{S-Conclusion} 
In this study, we present the observations of an EUV wave on 28 October 2021, which triggered  oscillations in several nearby loop systems. The wave was connected with  a filament eruption and a halo CME. The EUV wave consists of wave and non-wave components, wave trains, and several stationary fronts. The observation of fast-mode components, wave trains, and loop oscillations support the idea of the wave nature of this kind of EUV wave events \citep[e.g.][]{Thompson1998, Wang2000, Vrsnak2002, Pomoell2008, Schmidt2010}, while the non-wave and stationary fronts support the proposed non-wave model \citep{Delannee1999, Delannee2008}. Moreover, both these features can be explained by the hybrid model \citep{Chen2002,Chen2005}. The observed EUV wave trains are explained by the fluctuations of the magnetic field lines  seen by 
EUV loop oscillations.
We derive the coronal magnetic field, loop density ratios, and other parameters of the oscillating loops. The loop density ratios are derived using both DEM analysis and theoretical model. The values calculated from both methods are consistent. Furthermore, we find that the EUV wave propagates ahead of the leading edge of the CME using STEREO-A and LASCO data sets. This supports the earlier findings and models that show an EUV wave can be driven by a CME \citep{Ma2011, Patsourakos2009, Patsourakos2012, Cunha-Silva2015, Cunha-Silva2018}.

\begin{acks}
We would like to thank the referee for the constructive comments and suggestions.
The authors thank the open data policy of SDO, STEREO and SOHO instruments.  We would like to thank Aaron Peat for reading the manuscript to improve the language. We thank the editor for helping during the editorial process. We made use of NASA's Astrophysics Data System Bibliographic Services.
\end{acks}

\begin{authorcontribution}
PD did the data analysis and wrote the draft of the paper. RC, AKA, BS, and RJ wrote the substantial parts of the manuscript and contributed to the interpretation.  All the authors did a careful proofreading of the text and references.
\end{authorcontribution}

\begin{funding}
{PD thanks to the CSIR, New Delhi for the research fellowship. This research is supported by the Research Council of Norway through its Centres of Excellence scheme, project number 262622.} 
\end{funding}

\begin{dat}
The data sets analyzed during the current study are available at
\url{http://jsoc.stanford.edu/},  \url{https://stereo-ssc.nascom.nasa.gov/data.shtml}, and
\url{https://cdaw.gsfc.nasa.gov/}.
\end{dat}

\begin{conflict}
The authors declare that they have no conflicts of interest.
\end{conflict}

  
\bibliographystyle{spr-mp-sola1}
\bibliography{references}  

     
\end{article} 
\end{document}